%% file: main.tex
\documentclass[lettersize,journal]{IEEEtran}

\input{preamble_BM}
\makeatletter
\let\old@ps@headings\ps@headings
\let\old@ps@IEEEtitlepagestyle\ps@IEEEtitlepagestyle
\def\psccfooter#1{%
    \def\ps@headings{%
        \old@ps@headings%
        \def\@oddfoot{\strut\hfill#1\hfill\strut}%
        \def\@evenfoot{\strut\hfill#1\hfill\strut}%
    }%
    \def\ps@IEEEtitlepagestyle{%
        \old@ps@IEEEtitlepagestyle%
        \def\@oddfoot{\strut\hfill#1\hfill\strut}%
        \def\@evenfoot{\strut\hfill#1\hfill\strut}%
    }%
    \ps@headings%
}
\makeatother

\begin{document}

\title{Network Topology Reconfiguration: Optimal Transition Planning}

\author{
\IEEEauthorblockN{Basel Morsy$\dagger$*, Jochen Stiasny*, Adolfo Anta$\dagger$, Jochen Cremer$\dagger$*\\}
\IEEEauthorblockA{ $\dagger$ Austrian Institute of Technology AIT GmbH, Vienna, Austria\\
$^*$TU Delft, Delft, The Netherlands\\
B.Morsy@tudelft.nl
}
}

\maketitle

\bstctlcite{IEEEexample:BSTcontrol}

\begin{abstract}
Network topology reconfiguration (NTR) can reduce power system operating costs by co-optimizing generation dispatch and substation switching, but the reported savings describe a target operating point rather than a way to reach it. Reaching this operating point requires a sequence of intermediate operating points, each satisfying the AC power flow equations and thermal limits; because each topology admits its own feasible dispatch region, a naive transition that switches or redispatches first can drive intermediate flows past their thermal limits. Existing methods leave this gap open: snapshot NTR identifies a target but not a route. We formulate the Optimal Transition Planning (OTP) problem, co-optimizing the switching sequence and dispatch trajectory subject to AC feasibility at every intermediate point. We solve this problem with a receding-horizon framework: a DC planner proposes a trajectory that is certified against an AC feasibility filter, and infeasible topologies are excluded using reusable combinatorial cuts. Case studies on congested PGLib-OPF systems up to 1354-bus show that the method produces AC-feasible transitions that reduce operating cost by up to $18.4\%$ compared to the no-switching ACOPF solution on commodity hardware.
\end{abstract}



\input{Sections/Introduction}

\input{Sections/MotivatingExample}

\input{Sections/ProblemFormulation}

\input{Sections/Method}

\input{Sections/Results}

\input{Sections/Discussion}

\input{Sections/Conclusion}

\appendix
\input{Sections/Appendix}


\bibliographystyle{IEEEtran}
\bibliography{ref.bib}
\end{document}

%% file: preamble_BM.tex
\usepackage[cmex10]{amsmath}
\usepackage{array}
\usepackage{mathtools}
\usepackage{copyrightbox}
\usepackage{amssymb}
\usepackage{amsfonts}
\usepackage{mathtools}
\usepackage{cmap}
\usepackage{bm}
\usepackage{graphics}
\usepackage[T1]{fontenc}
\usepackage[obeyDraft]{todonotes}
\usepackage{siunitx}
\usepackage{tabularx,booktabs}
\newcommand{\cmark}{\checkmark}
\newcommand{\xmark}{\ensuremath{\times}}
\usepackage{cite}
\usepackage[hidelinks]{hyperref}
\usepackage{doi}
\usepackage{setspace}
\definecolor{navyblue}{rgb}{0.0, 0.0, 0.5}
\usepackage{graphicx}

\usepackage[export]{adjustbox}
\usepackage{float}
\usepackage{placeins}
\usepackage{nomencl}
\usepackage{enumitem}
\usepackage{algorithm, algorithmic}

\usepackage{makecell}
\usepackage{multirow}
\usepackage{subcaption}
\usepackage[font=footnotesize]{caption}
\usepackage{svg}

\makenomenclature
\newfloat{model}{htbp}{lop}
\floatname{model}{Model}

\renewcommand{\nomgroup}[1]{%
\ifthenelse{\equal{#1}{M}}{\item[\textbf{Models}]}{%
\ifthenelse{\equal{#1}{S}}{\item[\textbf{Sets and Indices}]}{%
\ifthenelse{\equal{#1}{V}}{\item[\textbf{Variables}]}{%
\ifthenelse{\equal{#1}{P}}{\item[\textbf{Parameters}]}{}}}}}

\usepackage{pgfplots}
\pgfplotsset{compat=1.18}

\definecolor{barblue}{HTML}{4682B4}
\definecolor{barred}{HTML}{CC3333}

\usepackage{tikz}
\usetikzlibrary{shapes.misc}
\usetikzlibrary{shapes.geometric, arrows}

\tikzstyle{startstop} = [ellipse, minimum width=3cm, minimum height=1cm,text centered, draw=black]
\tikzstyle{process} = [rectangle, minimum width=4cm, minimum height=1cm, text centered, draw=black, fill=blue!20]
\tikzstyle{decision} = [diamond, minimum width=3cm, minimum height=1cm, text centered, draw=black, fill=red!30]
\tikzstyle{init} = [rectangle, minimum width=4cm, minimum height=2cm, text centered, draw=black, fill=green!20]
\tikzstyle{arrow} = [thick,->,>=stealth]
\makeatletter
\g@addto@macro\normalsize{%
  \setlength{\abovedisplayskip}{5pt plus 2pt minus 2pt}%
  \setlength{\belowdisplayskip}{5pt plus 2pt minus 2pt}%
  \setlength{\abovedisplayshortskip}{2pt plus 2pt}%
  \setlength{\belowdisplayshortskip}{3pt plus 2pt}%
}
\makeatother

%% file: Sections/Introduction.tex
\section{Introduction}
\label{sec:intro}

Transmission networks have long been considered static assets, with most operational planning (aside from maintenance) assuming a fixed grid topology. This assumption underutilizes the grid's operational flexibility, leading to cost inefficiencies. To improve the grid's flexibility, operators can apply network topology reconfiguration (NTR). There is a substantial body of research on the utility of NTR: co-optimizing generation dispatch with topology yields cost savings~\cite{fisher_optimal_2008, hedman_co_optimization_2010}, relieves congestion~\cite{han_congestion_2015}, enhances voltage stability~\cite{wang_toward_2017}, reduces renewable curtailment~\cite{nasrolahpour_stochastic_2015}, aids post-contingency correction~\cite{morsy_security_2022, saavedra_day_ahead_2020}, and limits short-circuit currents~\cite{namchoat_optimal_2013}; comprehensive reviews are given in~\cite{aziz_review_2021,numan_role_2023}. These works focus on finding an optimal topology and dispatch at a single point in time. This constitutes a snapshot solution.

However, an optimal snapshot solution is not directly actionable, as the system cannot transition instantaneously from one operating point to another. Realizing NTR requires a well-defined \emph{transition plan} in which intermediate operating points respect steady-state feasibility and ramp-rate limits. The literature offers little tooling for constructing such planning. As a result, optimal topologies are often left unrealized or applied only partially based on operator experience, which limits the practical value of NTR.

Some work has been presented on studying the topology transition plan. In~\cite{9872531}, a transition-aware optimal transmission switching model was introduced that optimizes the sequence of switching actions from an initial to a target topology. However, the model assumes that the TSO knows a priori the number of transition steps required, considers only topology transitions rather than full operating-point transitions, and does not account for AC feasibility. In~\cite{liu_static_2012}, the security implications of the switching event itself were considered in a multi-period transmission switching setting, again without co-planning the generation trajectory. Reinforcement learning has also been applied to topology control~\cite{dorfer_power_2022, subramanian_exploring_2020}, but these target real-time operation rather than transition planning, control only the topology with dispatch resolved separately by the environment, and give no guarantee of intermediate-state feasibility.

Two adjacent lines of work inform the transition problem. First, model predictive control (MPC) for corrective control~\cite{Martin_2017} shares the sequential, ramp-constrained structure but excludes topology. Second, sequential switching in distribution restoration~\cite{Arif_2022} targets radial grids. Closest to our setting, \cite{BASTIANEL2026111527} solves a multi-period stochastic substation-switching model for hybrid AC/DC grids, but selects an independent topology per dispatch period rather than planning the trajectory between them.

Other prominent challenges in switching problems are tractability (solving the problem in a reasonable time) and AC-feasibility (respecting the AC physics of the grid). These challenges have been addressed from several angles: sensitivity-based heuristics~\cite{ruiz_tractable_2012, fuller_fast_2012, ruiz_fast_2011}, candidate pre-screening~\cite{liu_heuristic_2012, barrows_computationally_2012}, decomposition schemes~\cite{villumsen_column_2011, Nasrolahpour_2012, Heidarifar_2021}, tighter formulations and bound strengthening~\cite{fattahi_bound_2019, ruiz_security_constrained_2016}, and, more recently, learning-based methods~\cite{han_learning_based_2022, bugaje_real_time_2023}. These same challenges are inherent to planning transition trajectories.

\begin{table*}[h]
\centering
\caption{Positioning of representative prior work on topology reconfiguration and transition planning; the \emph{technical} constraints refer to ramp rates and switching budget, and \emph{physical} constraints refer to AC feasibility in intermediate operating points. \cmark: modeled; \xmark: not modeled.}
\label{tab:litreview}
\footnotesize
\setlength{\tabcolsep}{6pt}
\begin{tabular}{l cc cc cc}
\toprule
 & \multicolumn{2}{c}{Power flow model} & \multicolumn{2}{c}{Problem structure} & \multicolumn{2}{c}{Transition constraints} \\
\cmidrule(lr){2-3}\cmidrule(lr){4-5}\cmidrule(lr){6-7}
Reference & DC & AC & Sequential & Snapshot & Technical & Physical \\
\midrule
\cite{fisher_optimal_2008, hedman_co_optimization_2010, han_congestion_2015} & \cmark & \xmark & \xmark & \cmark & \xmark & \xmark \\
\cite{BASTIANEL2026111527}            & \xmark & \cmark & \xmark & \cmark & \xmark & \xmark \\
\cite{Nasrolahpour_2012, Heidarifar_2021}      & \xmark & \cmark & \xmark & \cmark & \xmark & \xmark \\
\cite{9872531, liu_static_2012}            & \cmark & \xmark & \cmark & \xmark & \xmark & \cmark \\
\cite{dorfer_power_2022, subramanian_exploring_2020} & \xmark & \cmark & \cmark & \xmark & \xmark & \xmark \\
\midrule
\textbf{This work}                                                    & \xmark & \cmark & \cmark & \xmark & \cmark & \cmark \\
\bottomrule
\end{tabular}
\end{table*}

Table~\ref{tab:litreview} is a comparison of the literature to our work based on three criteria: the power flow fidelity (DC vs.\ AC), the problem structure (a single \emph{snapshot} optimization vs.\ a \emph{sequential} decision over a horizon), and whether the two faces of a transition are jointly enforced: the \emph{technical} transition constraints (generator ramp limits and per-step switching budgets) and the \emph{physical} transition constraints (power flow feasibility of every intermediate operating point). To the best of our knowledge, no prior method co-plans an AC-feasible operating-point trajectory \emph{and} a physically realizable switching sequence within a single sequential framework. This is the gap that this work addresses.

The main contributions of this paper are as follows:
\begin{itemize}

    \item We formulate the Optimal Transition Planning (OTP) problem, which co-plans the substation switching sequence and the generation dispatch trajectory subject to AC feasibility and the technical transition limits at every intermediate operating point.
        
    \item We propose a receding-horizon (MPC) framework that solves the OTP problem tractably.

    \item We ensure AC feasibility for each step in the trajectory through a topology-filtering scheme that excludes AC-infeasible topologies with combinatorial no-good cuts.
        
\end{itemize}

We demonstrate this methodology on systems up to 1354-bus. The rest of the paper is structured as follows: Section \ref{sec:motivation} illustrates the need for transition planning, Section \ref{sec:formulation} formalizes the problem, Section \ref{sec:method} presents the solution approach, Section \ref{sec:results} presents the case studies, and lastly, discussion and conclusion in sections \ref{sec:discussion} and \ref{sec:conclusion} respectively.

%% file: Sections/MotivatingExample.tex
\section{Motivating Example: The Transition Dilemma}
\label{sec:motivation}

\begin{figure*}[!t]
    \centering
    \includegraphics[width=0.9\textwidth]{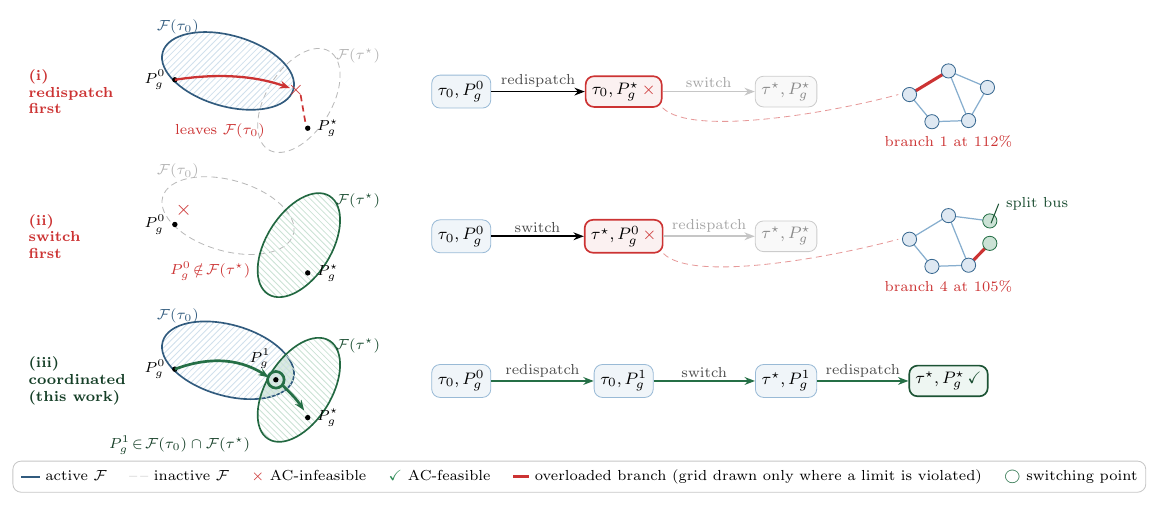}
    \caption{The transition dilemma on the IEEE 5-bus system. The current operating point $(\tau_0,P_g^0)$ and the NTR-optimal target $(\tau^\star,P_g^\star)$ differ in both topology and dispatch, and only one action type is executed per step. Left: dispatch space, where each topology induces its own feasible region $\mathcal{F}(\tau)$, solid when active and dashed when inactive. Right: the resulting sequence of operating points, with the network drawn only where a thermal limit is violated. Rows~(i) and~(ii) are the two ad-hoc orderings, each passing through an AC-infeasible state (greyed); row~(iii) is the coordinated transition, through an intermediate dispatch $P_g^1$ feasible under both topologies. Loadings use the full AC power flow.}
    \label{fig:motivating}
\end{figure*}

Network topology reconfiguration (NTR) as a snapshot co-optimizes generator dispatch and substation switching to identify a cheaper operating point, but it returns a static target: it prescribes \emph{where} the system should be, not \emph{how to get there} from the current state. Since this transition needs to be executed step by step, we restrict each step to a single action, either a generation redispatch or a switching action, but not both, so that every intermediate operating point is unambiguous and each move can be confirmed before the next is issued. Figure~\ref{fig:motivating} examines what this restriction implies on the IEEE 5-bus system, where the initial operating point consisting of topology $\tau_0$ and dispatch $\bm{P}_g^0$ and the NTR-optimal operating point $(\tau^\star, \bm{P}_g^\star)$ differ in both topology and dispatch. All flows reported below are computed with the full AC power flow model.

Under the single-action restriction, the shortest transition can be realized in exactly two ways, shown as rows~(i) and~(ii) of Figure~\ref{fig:motivating}. Redispatching first applies $\bm{P}_g^\star$ to the unmodified topology $\tau_0$. That dispatch was optimized for the reconfigured network and relies on the substation split to redistribute flows; without it, branch~1 carries $112\%$ of its thermal rating. Switching first applies $\tau^\star$ while holding $\bm{P}_g^0$ constant. The split changes the bus admittance matrix and pushes power through corridors the initial dispatch was never designed to accommodate, and branch~4 reaches $105\%$. Both intermediate states violate thermal limits, so neither ad-hoc ordering is executable.

The reason is geometric, as the left column of Figure~\ref{fig:motivating} shows. Each topology induces its own feasible dispatch region $\mathcal{F}(\tau)$, the set of generator setpoints that satisfy the AC power flow equations and the operating limits under that topology. Two topologies differ in which buses are electrically connected and therefore in their admittance matrices, so $\mathcal{F}(\tau_0) \neq \mathcal{F}(\tau^\star)$, and in this example $\bm{P}_g^\star \notin \mathcal{F}(\tau_0)$ and $\bm{P}_g^0 \notin \mathcal{F}(\tau^\star)$. Path~(i) drives the operating point out of $\mathcal{F}(\tau_0)$ while $\tau_0$ is still the active topology; path~(ii) activates $\mathcal{F}(\tau^\star)$ at a dispatch that lies outside it. The infeasibility thus originates in the AC physics and the thermal limits, not in the ramp rates or switching budgets imposed later in this work, and no relaxation of those transition parameters can remove it.

Row~(iii) shows the resolution. A coordinated transition first redispatches within $\mathcal{F}(\tau_0)$ to an intermediate setpoint $\bm{P}_g^1 \in \mathcal{F}(\tau_0) \cap \mathcal{F}(\tau^\star)$, then switches at that point, where the same dispatch is feasible immediately before and after the switch, and only then redispatches to $\bm{P}_g^\star$. Every intermediate operating point satisfies the constraints of the topology active at its step, at the price of a longer sequence of actions.

The dispatch-space depiction is schematic in two respects. The regions are drawn as ellipses for clarity; in general, the feasible spaces are non-convex due to the AC power flow constraints. More importantly, the two regions need not intersect: when $\mathcal{F}(\tau_0) \cap \mathcal{F}(\tau^\star) = \emptyset$, no single intermediate dispatch bridges them, and the transition must pass through one or more intermediate topologies whose regions connect the two. A snapshot NTR solution may even be unreachable from the initial state; reachability analysis is beyond the scope of this work. Instead, we focus in the remainder of this paper on computing the interleaved sequence of switching and redispatch actions that keeps every intermediate operating point feasible. We refer to this sequence as the \emph{transition plan}.

%% file: Sections/ProblemFormulation.tex

\section{Problem Formulation}
\label{sec:formulation}

\begin{figure}
    \centering
    \includegraphics[width=1.0\linewidth]{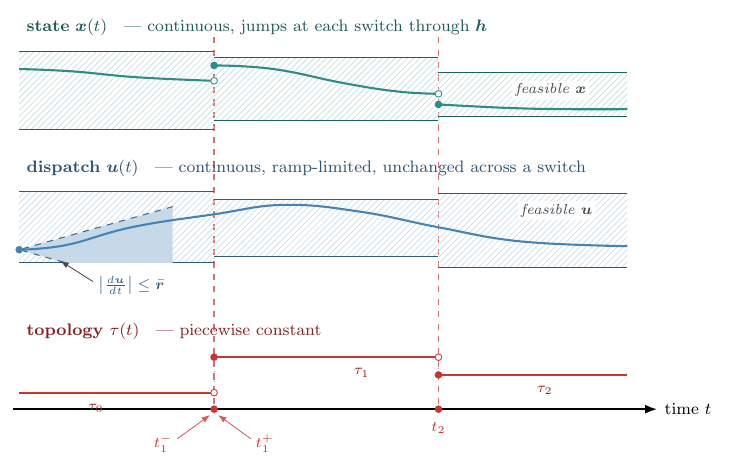}
    \caption{Evolution of state, dispatch, and topology over time scales.}
    \label{fig:timescales}
\end{figure}

We begin by defining the system and its variables; throughout, boldface denotes real-valued vectors and vector-valued maps, while the topology $\tau$ and the binary switch variables are written in plain italic. The system state $\bm{x}(t)$ evolves continuously in time and captures all electrical quantities that characterize the operating point (e.g., voltage magnitudes, phase angles, and branch flows). The operator controls two distinct types of inputs: continuous control variables $\bm{u}(t)$, representing generator dispatch setpoints that can be adjusted smoothly over time, and discrete control variables $\tau$, representing switching actions (e.g., breaker operations in substations) that change the network topology at discrete instants. Figure~\ref{fig:timescales} illustrates these different variables across time.
The switching actions are collected in the set $\mathcal{K}$ and indexed by $k$; switch $k$ occurs at an instant $t_k$ with left and right limits $\tau(t_k^{-})$ and $\tau(t_k^{+})$, so that $\tau(t_k^{+}) - \tau(t_k^{-})$ is the topology change corresponding to the $k$-th switching action.

\begin{model}[htbp]
\caption{Generic Optimal Transition Planning Model}
\label{model:gtp}
\begin{subequations} \label{eq:GTP}
    \begin{align}
        \begin{split}\label{eq:obj}
            \min_{\bm{x}, \bm{u}, \tau} \;\; & \int_0^{T} w(t)\big[\,C(\bm{u}(t))+C(\bm{x}(t))\,\big]\,dt \\ &+ \sum_{k \in \mathcal{K}} C\big(\tau(t_k^{+}) - \tau(t_k^{-})\big) \quad + \iota\, C^{\star}\!\big(\bm{x}(T),\bm{x}^{\star}\big)
        \end{split}\\        
        \textrm{s.t.:} \quad & \bm{g}\big(\bm{x},\bm{u}, \tau, \tfrac{d\bm{u}}{dt},\Delta\tau\big) \leq 0 \label{eq:inequality}\\
        & \bm{h}(\bm{x},\bm{u}, \tau)=0 \label{eq:equality}
    \end{align}
\end{subequations} 
\end{model}

We formulate the optimal transition planning problem (OTP) as an optimal control problem over a finite time budget $T$ in Model~\ref{model:gtp}. Where $\bm{h}$ is the AC power flow equations, $\bm{g}$ specifies the technical limits and transition constraints, and $C$ determines the appropriate cost structure to the application. The result is a transition plan that co-optimizes topology and dispatch, ensuring that every intermediate operating point satisfies both the power flow equations and the operational limits in effect at that step.

The objective~\eqref{eq:obj} minimizes the total transition cost. The integral term is the cost of continuous operation, combining the dispatch cost $C(\bm{u}(t))$ (e.g., generation cost from an OPF model) with an optional state-preference cost $C(\bm{x}(t))$ (for instance, proximity to operating limits or losses), weighted across time by $w(t)$. The summation term charges each switching action the cost $C(\tau(t_k^{+}) - \tau(t_k^{-}))$ of the topology change it makes, summed over the set $\mathcal{K}$ of switching actions.

The final term is an optional destination penalty $C^{\star}$, activated by $\iota = 1$ when a target operating point $\bm{x}^{\star}$ (for example, a snapshot NTR solution) is prescribed, and dropped ($\iota = 0$) otherwise. Throughout this work $\iota = 0$, so the planner may reach any cost-reducing operating point feasible from the initial state; the target-guided case is a special case. The motivating example of Section~\ref{sec:motivation} is the exception, using a fixed NTR-optimal target to illustrate the transition difficulty most clearly.

The inequality constraints~\eqref{eq:inequality} encode both the technical limits of the system and the transition constraints that govern how the controls evolve. The dependence on the continuous derivative $d\bm{u} / dt$ captures ramp rate limits on generator output, while the dependence on the discrete difference $\Delta \tau$ captures switching budgets that limit the number of topology changes per step. The equality constraints~\eqref{eq:equality} enforce the physical laws governing the power system, namely the power flow equations that link the state $\bm{x}$ to the controls $\bm{u}$ and $\tau$ at every instant.

The state $\bm{x}$ is determined by the two controls through the power flow $\bm{h}(\bm{x},\bm{u},\tau) = 0$, so the object we plan is the joint control trajectory $(\bm{u}(\cdot), \tau(\cdot))$. Discretizing time into steps $t = 0, 1, 2, \dots$, with $t=0$ the initial operating point, the integral in~\eqref{eq:obj} becomes a finite summation, the continuous derivative $d\bm{u}/dt$ is replaced by a forward difference $\Delta \bm{u}_t = \bm{u}_{t+1} - \bm{u}_t$, and the constraints are enforced at each discrete step. The transition plan is then the sequence of dispatch--topology pairs $\{(\bm{u}_t, \tau_t)\}$, which steers the system from its initial operating point to the target configuration while satisfying all constraints at every intermediate step; its length is not fixed in advance.

Two features of Model~\ref{model:gtp} make it impractical to solve as posed. First, the time budget $T$ is not known a priori, and the number of discrete steps needed to realize the transition is itself part of what we seek rather than a given input. Second, even if the horizon were fixed, Model~\ref{model:gtp} is a single multi-period mixed-integer program that couples the continuous dispatch and the discrete switching decisions across every step, a structure closely related to the unit commitment problem. For a transition that may span many steps, solving this monolithic problem to optimality is computationally intractable.

%% file: Sections/Method.tex

\section{Methodology}
\label{sec:method}

The methodology developed in this section constructs a transition plan, the realized control trajectory $\{(\hat{\bm{u}}_t, \hat{\tau}_t)\}$, that improves the operating cost while keeping every intermediate operating point AC-feasible and within the operational limits.

\begin{figure*}[ht]
  \centering
  \includegraphics[width=\textwidth]{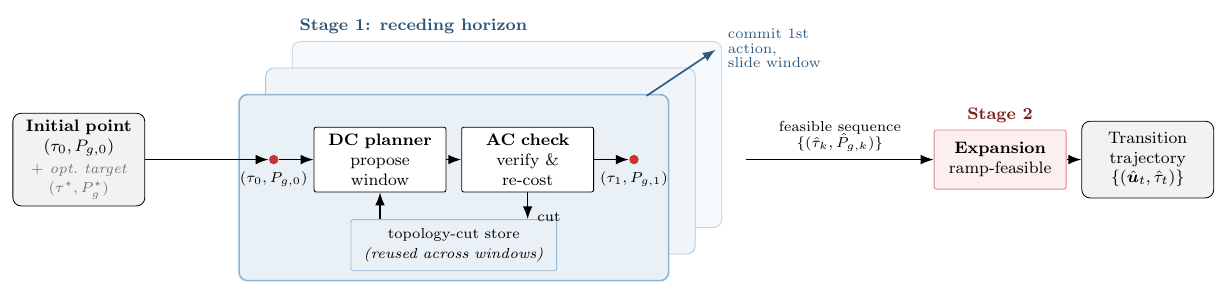}
  \caption{Two-stage overview of the methodology. \textbf{Stage~1} is a receding horizon over look-ahead windows (the stacked cards). The front window runs an \emph{inner loop}: from an initial point, the DC planner proposes a window trajectory and the AC check verifies it and returns its corrected cost, feeding cuts and cost back until the window is AC-feasible and improving; committing its first action and sliding forward yields the next window (\emph{outer loop}). The resulting feasible topology sequence is passed to \textbf{Stage~2}, which expands it into the ramp-feasible transition trajectory $\{(\hat{\bm{u}}_t, \hat{\tau}_t)\}$, AC-feasible at every intermediate step.}
  \label{fig:method_overview}
\end{figure*}
We obtain this trajectory in two stages (Figure~\ref{fig:method_overview}). \textbf{Stage~1} searches for a \emph{feasible topology sequence} with the receding-horizon scheme of Section~\ref{subsec:mpc}, in which a DC planner and an AC check are iterated to plan a window whose first action is then committed as the window slides forward until convergence. \textbf{Stage~2} \emph{expands} the committed sequence into a ramp-feasible dispatch trajectory (Section~\ref{subsec:expansion}). The result is AC-feasible at every step, obeys the operational constraints, and improves the objective.

Dispatch and topology share a single discretized (physical) time grid indexed by $t$, on which the planner imposes a \emph{single-action rule}: each step either redispatches or switches, but not both. A switching step, therefore, holds the dispatch fixed.

\subsection{Receding-horizon scheme}
\label{subsec:mpc}

The number of steps needed to reach a reconfigured operating point is not known a priori, and solving Model~\ref{model:gtp} over the full horizon is an intractable mixed-integer nonlinear program. Stage~1 therefore solves it in a receding horizon with two nested loops (Figure~\ref{fig:method_overview}).

The \emph{outer loop} advances the transition. At each iteration, we plan over a fixed look-ahead window of $H$ steps, commit only the first action, advance the state to the resulting operating point, and slide the window forward. Iterating steers the system toward a lower-cost configuration one action at a time, requires no a priori horizon, and, by re-planning from the realized state at every step, remains robust to the gap between the planner and the physical grid. The outer loop terminates when the committed topology no longer changes and the terminal-cost improvement falls below a tolerance $\epsilon$, indicating that the planner has settled at a local optimum, or when an optional target operating point is reached.

The \emph{inner loop} solves a single window by alternating the two components of Stage~1. The DC planner (Section~\ref{subsec:mpcx}, Model~\ref{model:mpcdc}) proposes the least-cost window trajectory under DC physics. The AC check (Section~\ref{subsec:filter}, Model~\ref{model:mpacopf}) verifies that every intermediate operating point of that trajectory is AC-feasible and that the terminal cost is improving. If successful, it returns the trajectory to the outer loop. If the check is unsuccessful, we add cuts and re-solve the DC planner.

Both models are instantiations of the generic model (Model~\ref{model:gtp}): we introduce each by stating what its state $\bm{x}$, controls $(\bm{u},\tau)$, power flow $\bm{h}$, and cost $C$ stand for, so the reader can read the concrete model against the generic template.

\subsection{Planning model: MPC-DC}
\label{subsec:mpcx}

The planner instantiates Model~\ref{model:gtp} with DC physics, giving the mixed-integer linear program of Model~\ref{model:mpcdc}. The continuous control is the generator dispatch, $\bm{u} = \bm{P}$, priced by the terminal stage cost $C(\bm{u}) = \sum_{g} C_g P_{g,H}$; the discrete control is the topology, $\tau = (z^r, z^c)$, where $z^r$ and $z^c$ are the reconfiguration and coupler binary variables respectively; and the state reduces to the bus voltage angles, $\bm{x} = \bm{\delta}$. The equality $\bm{h}$ is the DC power flow~\eqref{eq:DCPF} (Appendix~\ref{sec:pf}) together with the Augmented Network Representation (ANR)~\cite{morsy_security_2022} for substation switching (constraints~\eqref{eq:sss}, Figure~\ref{fig:ANR}, Appendix~\ref{sec:switching}); the inequality $\bm{g}$ gathers the DC generation and thermal limits with the per-step switch budgets $\lambda_r,\lambda_c$. Every variable carries the step index $t$ and is evaluated on the corresponding network topology $\tau_t$; the initial step $t=0$ is fixed to the current operating point $(\bm{P}_0, \tau_0)$, and a check rejects any integer solution whose network is disconnected.

\begin{model}[htbp]
\caption{MPC-DC: DC instantiation of Model~\ref{model:gtp} over a window}
\label{model:mpcdc}
\begin{subequations} \label{eq:MPCDC}
\begin{align}
    \min_{\bm{P},\bm{\delta},z} \;\; & \sum_{g \in \mathcal{G}} C_{g}\,P_{g,H} \label{eq:mpcdc_obj}\\
    \textrm{s.t.:} \quad
    & \text{DC power flow~\eqref{eq:DCPF}, ANR switching~\eqref{eq:sss}} \label{eq:mpcdc_h}\\
    & \text{DC generation and thermal limits} \label{eq:mpcdc_g}\\
    & \xi_{t} \in \{0,1\} \label{eq:trans_x}\\
    & \big\lVert z^{r}_{t}-z^{r}_{t-1}\big\rVert_1 \le \lambda_{r} \xi_{t}, \ \ \big\lVert z^{c}_{t}-z^{c}_{t-1}\big\rVert_1 \le \lambda_c \xi_{t} \label{eq:trans_budget}\\
    & \xi_{t} = 1 \;\Rightarrow\; P_{g,t} = P_{g,t-1}, \quad \forall g \in \mathcal{G} \label{eq:trans_freeze}\\
    & (\bm{P}_0,z_0)\ \text{fixed}, \quad t \in \{1,\dots,H\} \label{eq:mpcdc_ic}
\end{align}
\end{subequations}
\end{model}

The objective~\eqref{eq:mpcdc_obj} is the terminal dispatch cost $C(\bm{u})$; \eqref{eq:mpcdc_h} and~\eqref{eq:mpcdc_g} instantiate the power flow $\bm{h}$ and the limits $\bm{g}$ of Model~\ref{model:gtp}. The selector $\xi_t$ enforces the single-action rule: the budget~\eqref{eq:trans_budget} permits a switch change only when $\xi_t = 1$, so a redispatch step ($\xi_t = 0$) moves no switch, and~\eqref{eq:trans_freeze} freezes the dispatch across a switching step; \eqref{eq:mpcdc_ic} fixes the initial step. The freeze is imposed exactly here, in the planner; because an exact freeze is generally infeasible under AC physics, the multi-period ACOPF that certifies the plan relaxes it to a soft penalty (Section~\ref{subsec:filter}). The detailed switch budget belongs to the switching model of Appendix~\ref{sec:switching}, and the per-step ramp limit is omitted here and restored by expansion (Section~\ref{subsec:expansion}).

We warm start each MPC window with the last known feasible trajectory. In the first window, this is the stay-at-current-operation plan, in which no switch moves and the dispatch is held, which is integer-feasible by construction. In subsequent windows, we supply the planned but uncommitted feasible trajectory from the previous window. To bound the computational budget, we impose a per-window cap on the number of DC re-solves of the inner loop.

\subsection{AC feasibility check}
\label{subsec:filter}

The AC check instantiates Model~\ref{model:gtp} with AC physics: the state is the full AC operating point $\bm{x} = (\bm{V}, \bm{\delta}, \bm{P}, \bm{Q})$, the equality $\bm{h}$ is the AC power flow~\eqref{eq:ACPF} (Appendix~\ref{sec:pf}), and the dispatch $\bm{u} = \bm{P}$ is free. In contrast to the planner, the discrete control $\tau$ is not optimized here: it is fixed to the planned switch pattern $\tau_t$, so the check verifies the feasibility of a given topology sequence rather than searching over topologies.

The DC plan is certified against AC physics in two stages. First, we identify whether the chosen topologies are individually AC-feasible using a single-period topology proxy. Second, we check for AC-infeasible topology sequences in a multi-period verification model.

\emph{Single-period topology proxy.} Each unique topology $\tau_t$ in the planned sequence is screened individually. We solve a single-period ACOPF on the AC power flow~\eqref{eq:ACPF} in which the switching binaries are relaxed to the continuous interval $[0,1]$, with the discrete switch behavior represented through the big-M switching constraints~\eqref{eq:sss}, and the objective replaced by a proximity term $\min \lVert z - \hat{z}_t \rVert_2^2$ that drives the relaxed switches toward $\hat{z}_t$, the target switch pattern of the planned topology $\tau_t$. If the recovered switch values match the target to within a drift tolerance, the topology admits an AC operating point and is accepted; otherwise it is declared AC-infeasible and excluded by a cut, and the planner is re-solved.

\emph{Multi-period verification.} A topology that is individually feasible may still be unreachable within a coordinated sequence. We therefore verify the full trajectory with the multi-period ACOPF of Model~\ref{model:mpacopf}, which enforces the AC power flow equations~\eqref{eq:ACPF} at every step on the network induced by $\tau_t$ and minimizes the terminal cost. The planner freezes the dispatch exactly across each switching step, but an exact freeze $\bm{P}_{t} = \bm{P}_{t-1}$ across the two topologies is generally infeasible under AC physics, because their losses differ. The verification therefore relaxes the freeze to a quadratic penalty, weighted by $\gamma$, on the dispatch jump $\bm{P}_{t} - \bm{P}_{t-1}$ at each switching step $t \in \mathcal{J} = \{t : \tau_t \neq \tau_{t-1}\}$. The penalty \emph{discourages but does not forbid} movement; its weight $\gamma$ balances the dispatch jump size against the terminal cost. Raising $\gamma$ shrinks the dispatch jump at the cost of deviating from the least cost dispatch trajectory. The dispatch jump requires a dynamic response and needs to be judged under the corresponding criteria.

\begin{model}[htbp]
\caption{MP-ACOPF: AC instantiation of Model~\ref{model:gtp} over the planned sequence}
\label{model:mpacopf}
\begin{subequations} \label{eq:MPACOPF}
\begin{align}
    \min_{\bm{P},\bm{Q},\bm{V},\bm{\delta}} \;\; & \sum_{g \in \mathcal{G}} C_{g}\,P_{g,H} + \gamma \sum_{t \in \mathcal{J}} \big\lVert \bm{P}_{t}-\bm{P}_{t-1} \big\rVert^2 \label{eq:mpacopf_obj}\\
    \textrm{s.t.:} \quad
    & \text{AC power flow~\eqref{eq:ACPF} on } \tau_t \label{eq:mpacopf_h}\\
    & \text{AC generation, thermal, voltage \& angle limits} \label{eq:mpacopf_g}\\
    & \text{topology fixed to the planned}\ \{\tau_t\} \label{eq:mpacopf_fix}
\end{align}
\end{subequations}
\end{model}

The outcome of the multi-period ACOPF determines the feedback to the planner. If it is infeasible, a \emph{partial-trajectory cut} is added for every consecutive topology pair of the planned sequence: for each adjacent pair $(\tau_t, \tau_{t+1})$ in the plan, the planner is thereafter forbidden from placing that pair at any two consecutive steps. If it is feasible but its terminal cost does not improve on the best certified plan, the whole sequence is forbidden by a \emph{full-trajectory cut}. If it is feasible and improving, the plan is accepted: its first action is committed and its topologies are recorded as AC-feasible.

All exclusions take the form of no-good cuts on the switching binaries ($z^r$, $z^c$), expressed on the switch vector $\bm{z} \in \{0,1\}^{|z^r|+|z^c|}$. A topology is excluded by the Hamming no-good cut
\begin{equation}\label{eq:nogood}
    \mathcal{H}(\bm{z}_t, \hat{\bm{z}}) = \sum_{j:\, \hat{z}_j = 0} z_{j,t} + \sum_{j:\, \hat{z}_j = 1} \big(1 - z_{j,t}\big) \;\geq\; 1,
\end{equation}
which is linear in the switch variables. Four families of cuts are maintained in a topology store and re-injected into every planner solve: connectivity cuts, added a check for any disconnected resulting network; AC-infeasibility cuts from the proxy; partial-trajectory cuts, one per consecutive topology pair of an AC-infeasible sequence, each forbidding that pair at any two consecutive steps; and full-trajectory cuts forbidding an entire non-improving sequence.

\emph{Relation to prior decomposition schemes.} The structure just described --- a relaxed master that proposes topologies, an AC subproblem that certifies them, and cuts that feed infeasibility back --- is the multi-period transition-planning generalization of the two-level \emph{snapshot} schemes of~\cite{Heidarifar_2021} (integer no-good cuts) and~\cite{Nasrolahpour_2012} (Benders cuts). Our filter uses combinatorial no-good cuts. An alternative, following~\cite{Nasrolahpour_2012}, fixes the planned switches and dispatch in the AC check and reads its multipliers to form generalized-Benders cuts that bound an auxiliary master cost; because the AC subproblem is non-convex, these duals are only locally valid, so the cuts are heuristic and may exclude the true optimum. We nonetheless implement this Benders-cut variant as a literature-derived baseline (Appendix~\ref{sec:benders}) and compare it against our combinatorial cuts in Section~\ref{subsec:cut_comparison}.

\subsection{AC expansion}
\label{subsec:expansion}

The receding-horizon loop returns a committed trajectory of operating points in which the ramp limit was not enforced. Expansion restores ramp feasibility without altering any topology decision. Along the committed trajectory, switching steps are passed through unchanged: the dispatch adjustment they carry is the AC-feasible value fixed by the multi-period rectification and is treated as an atomic part of the switching action rather than a ramp-limited redispatch, so expansion leaves these steps intact. For a redispatch step that moves generation by more than one ramp budget, we compute the minimum number of intermediate steps
\begin{equation}\label{eq:expand_n}
    \kappa = \max_{g \in \mathcal{G}} \left\lceil \frac{\lvert P_{g}^{+} - P_{g}^{-} \rvert}{\alpha\, P_{g}^{\max}} \right\rceil ,
\end{equation}
where $P_{g}^{-}$ and $P_{g}^{+}$ are the dispatches at the two ends of the gap, and bridge the gap with a sequence of single-period ACOPFs that, at each step, minimize the dispatch distance $\lVert \bm{P}_t - \bm{P}^{+} \rVert$ to the gap endpoint subject to the per-step ramp limit $\lvert P_{g,t} - P_{g,t-1} \rvert \leq \alpha P_{g}^{\max}$. A gap that would require more than a preset number of intermediate steps is flagged rather than bridged. The result is a physically realizable trajectory in which every consecutive dispatch pair respects the ramp budget and every intermediate point remains AC-feasible.

%% file: Sections/Results.tex
\section{Case Studies}
\label{sec:results}

\subsection{Experimental setup}
All models are implemented in Julia, with the mixed-integer linear programs solved by Gurobi, and the non-linear programs by Ipopt. The test systems are the Congested Operating Conditions (API) variants of the PGLib-OPF benchmark library~\cite{pglib}, whose tighter line ratings and higher loading make reconfiguration economically meaningful. All experiments run on a laptop-class machine (Intel Core i7-1185G7 at $3.00$~GHz, $32$~GB RAM), so the reported solve times reflect commodity hardware.

\emph{Selecting splittable substations.} Allowing every substation to split is combinatorially prohibitive and largely wasteful, since reconfiguration helps only where the network is constrained. We therefore prescreen the splittable set: we solve a DC optimal power flow on the original network and designate as splittable the substations adjacent to lines at their thermal limit. Table~\ref{tab:main_results} reports the resulting count.

\emph{Hyperparameters.} Unless stated otherwise, the planner uses the settings in Table~\ref{tab:hyperparams}: a look-ahead horizon $H = 5$, per-step switching budgets $\lambda_r = 4$ and $\lambda_c = 1$, a ramp factor $\alpha = 0.1$ (at most $10\%$ of a generator's capacity per redispatch step), and a terminal-cost tolerance $\epsilon = 10^{-4}$.

\begin{table}[t]
\centering
\caption{Planner hyperparameters.}
\label{tab:hyperparams}
\footnotesize
\begin{tabular}{cll}
\toprule
Symbol & Meaning & Value \\
\midrule
$H$         & look-ahead horizon (steps)                  & $5$ \\
$\lambda_r$ & per-step reconfiguration budget             & $4$ \\
$\lambda_c$ & per-step coupler budget                     & $1$ \\
$\alpha$    & per-step ramp budget (fraction of $P_g^{\max}$) & $0.1$ \\
$\epsilon$  & terminal-cost convergence tolerance         & $10^{-4}$ \\
            & planner MILP time limit                     & $120$~s \\
            & AC proxy drift tolerance                    & $10^{-2}$ \\
            & maximum outer iterations                    & $20$ \\
            & max re-solve attempts per iteration         & $6$ \\
$\gamma$    & soft-freeze penalty weight                  & $10^{-3}$ \\
$\rho$      & Benders recourse weight (baseline)          & $1$ \\
\bottomrule
\end{tabular}
\end{table}

\emph{Comparators.} We compare the proposed method (combinatorial no-good cuts) against three alternatives: a \emph{Benders} variant using the generalized-Benders cuts of Appendix~\ref{sec:benders} (adapting~\cite{Nasrolahpour_2012}), a \emph{combined} variant that enables both cut types, and a \emph{naive DC$\to$AC} baseline that removes AC filtering entirely (Section~\ref{subsec:filter_necessity}). All four share the same planner, prescreen, systems, and time budgets, differing only in how the AC subproblem feeds back to the master.

\subsection{Numerical results}
Table~\ref{tab:main_results} reports the planned transition for each of the ten API systems, ordered by descending savings. The outcomes characterize the operating envelope of the method rather than a single average: reconfiguration pays only where the network is congested. Three systems (118, 793, 24) save $5\%$ or more, reaching $18.4\%$ on the 118-bus system; three others (30, 39, 57) save exactly zero, the planner declining to switch when no binding corridor rewards it; the remaining four show positive but sub-$5\%$ savings. The method thus recovers large savings where they exist and returns a short, near-trivial plan where they do not, which itself signals to the operator whether reconfiguration is worth executing.

\begin{table*}[t]
\centering
\caption{Transition-planning results on the PGLib-OPF API systems, ordered by descending savings and grouped by \emph{NTR potential}: high ($\geq 5\%$), low ($<5\%$), or none ($0\%$) savings. ``ACOPF'' is the initial ACOPF cost with no switching (the cost of doing nothing); savings is the reduction attained by the planned transition, given as an absolute value and as a percentage of the ACOPF cost. Steps lists the physical-trajectory length with its split into topology ($\#$T) and redispatch ($\#$D) actions. The last column is the savings attained by feeding the unfiltered DC topology plan directly into a multi-period ACOPF (\emph{infeasible} when no AC-realizable trajectory exists; a negative value means the naive plan is costlier than doing nothing).}
\label{tab:main_results}
\footnotesize
\setlength{\tabcolsep}{5pt}
\IfFileExists{Results/main_table.tex}{\input{Results/main_table.tex}}{\textit{\small [results table not yet generated]}}
\end{table*}

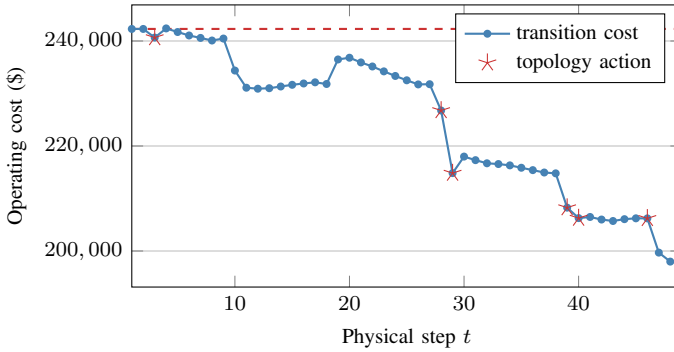
\begin{figure}[t]
    \centering
    \IfFileExists{Results/case_featured/cost_evolution.dat}{%
        \InputIfFileExists{Results/case_featured/meta.tex}{}{\def\CaseCzero{0}\def\CaseXmin{1}\def\CaseXmax{2}}%
        \begin{tikzpicture}
        \begin{axis}[
            width=\columnwidth, height=0.6\columnwidth,
            xlabel={Physical step $t$}, ylabel={Operating cost (\$)},
            ymajorgrids, scaled y ticks=false,
            yticklabel style={/pgf/number format/fixed, /pgf/number format/1000 sep={,}},
            legend pos=north east, legend cell align=left, font=\footnotesize,
            xmin=1, xmax=49
        ]
            \addplot[barred, dashed, thick, forget plot] coordinates {(\CaseXmin,\CaseCzero) (\CaseXmax,\CaseCzero)};
            \addplot[barblue, thick, mark=*, mark size=1.1pt]
                table[x=step, y=cost] {Results/case_featured/cost_evolution.dat};
            \addlegendentry{transition cost}
            \addplot[barred, only marks, mark=star, mark size=3.5pt]
                table[x=step, y=cost] {Results/case_featured/topo_steps.dat};
            \addlegendentry{topology action}
        \end{axis}
        \end{tikzpicture}
    }{\textit{\small [Figure data not yet available.]}}%
    \caption{Operating-cost trajectory of the physical (ramp-expanded) transition for the 118-bus API system. The dashed line is the initial ACOPF cost with no switching; stars mark the six switching steps and the remaining steps are ramp-feasible redispatch. Every intermediate point is AC-feasible.}
    \label{fig:cost_118}
\end{figure}

The planned transition is realized in two resolutions: the \emph{algorithmic} trajectory returned by the receding-horizon loop, with one committed action per iteration, and the \emph{physical} trajectory obtained after ramp expansion, which refines each committed redispatch into ramp-feasible sub-steps. For the 118-bus case this expands a compact eleven-step plan into $49$ physical steps. Figure~\ref{fig:cost_118} traces the physical trajectory: the operating cost descends from the initial ACOPF point to the reconfigured one, with the switching steps marked and every intermediate point AC-feasible.

\subsection{Necessity of AC feasibility filtering}
\label{subsec:filter_necessity}
The AC feasibility filter is the component that turns a DC plan into a realizable transition. To quantify its necessity, we compare against a naive baseline that removes it entirely: we run the DC planner alone, with no AC proxy, no multi-period verification, and no cuts, and feed the resulting topology sequence directly into a multi-period ACOPF. The last column of Table~\ref{tab:main_results} reports the outcome, expressed as savings for direct comparison with the filtered plan. The key observation is not that the naive plan is uniformly worse, since on several lightly loaded systems it coincides with the filtered plan, but that when it fails, it fails in ways an operator cannot absorb. On two of the ten systems (118 and 162) the unfiltered plan is not AC-realizable at all: the multi-period ACOPF over its topology sequence is infeasible, so the transition cannot be executed without violating the AC power flow or thermal limits at some intermediate step. One of these is the most valuable case, the 118-bus system, where the unfiltered plan would forfeit the entire $18.4\%$ saving by being unexecutable. On a third system (73) the naive plan is realizable but costlier than doing nothing ($-0.3\%$), converting a reconfiguration opportunity into a loss. This is the failure mode anticipated by the motivating example of Section~\ref{sec:motivation}, now observed at scale: the filter is inexpensive when the DC plan is AC-realizable and decisive when it is not, and the outcome cannot be predicted without applying it.

\subsection{Comparison of cut strategies}
\label{subsec:cut_comparison}
We compare the proposed combinatorial filter against the Benders-cut baseline and the combined ablation introduced above; Table~\ref{tab:cut_comparison} reports the savings and solve time of each configuration per system. Three patterns emerge.

First, the combinatorial filter is the most robust: on every system it returns an AC-feasible plan that never increases cost relative to the initial operating point, and on the most valuable case (118-bus) it recovers far more than either alternative ($18.4\%$ versus $7.1\%$). Second, the Benders baseline is competitive but brittle. On some systems it reaches a lower-cost terminal than the combinatorial filter, for instance $11.7\%$ versus $10.2\%$ on the 793-bus system and, together with the combined variant, $8.9$--$10.6\%$ versus $5.3\%$ on the 24-bus system, and it can converge in fewer planner re-solves; on others it terminates at a costlier local optimum, or commits no switching action at all (the 73-bus system), leaving the grid at its initial topology. This behavior follows from the heuristic caveat of Appendix~\ref{sec:benders}: the locally valid duals can steer the master toward a cheaper basin, but nothing prevents them from steering it into a feasible but suboptimal one, where the loop then settles. Third, the combined variant is inconsistent: on some systems the Benders cuts accelerate the combinatorial search, on others they degrade it below the combinatorial-only result, so enabling both does not reliably help. The combinatorial filter therefore does not dominate on cost; instead it certifies an AC-feasible plan on every system, which is the property that matters when the objective is a reliable transition rather than the smallest number of re-solves.

Taken together, these results corroborate the integer-cut design choice of the most recent snapshot method~\cite{Heidarifar_2021} and empirically expose the limitation of the dual-cut approach of~\cite{Nasrolahpour_2012} in the transition setting.

\begin{table*}[t]
\centering
\caption{Cut-strategy comparison on the PGLib-OPF API systems: savings (absolute and relative to the initial ACOPF cost) and wall-clock solve time for the proposed combinatorial filter, the Benders-cut baseline (Appendix~\ref{sec:benders}), and the combined ablation. The three systems that yield no savings under any strategy (30, 39, 57 in Table~\ref{tab:main_results}) are omitted. A savings of $0$ ($0.0\%$) marks a configuration that committed no switching action, so the terminal cost equals the initial ACOPF cost.}
\label{tab:cut_comparison}
\footnotesize
\setlength{\tabcolsep}{5pt}
\IfFileExists{Results/cut_comparison_table.tex}{\input{Results/cut_comparison_table.tex}}{\textit{\small [cut-comparison table not yet generated]}}
\end{table*}

%% file: Results/main_table.tex
\begin{tabular}{l l r r r r l r}
\toprule
NTR potential & System & Splittable substations & ACOPF (\$) & Savings (\$, \%) & Time & Steps ($\#$T/$\#$D) & Naive DC$\to$AC (\$, \%) \\
\midrule
\multirow{3}{*}{High} & 118  & 8  & \num{242296}  & \num{44683} (18.4\%) & 60 m 23 s & 49 (6/42) & \emph{infeasible} \\
                      & 793  & 17 & \num{58492}   & \num{5962} (10.2\%)  & 30 m 21 s & 25 (5/19) & \num{4127} (7.1\%) \\
                      & 24   & 5  & \num{115930}  & \num{6107} (5.3\%)   & 38.4 s    & 11 (2/8)  & \num{4683} (4.0\%) \\
\midrule
\multirow{4}{*}{Low}  & 5    & 1  & \num{76377}   & \num{2651} (3.5\%)   & 2.5 s     & 9 (2/6)   & \num{2651} (3.5\%) \\
                      & 73   & 9  & \num{352149}  & \num{3043} (0.9\%)   & 48 m 8 s  & 16 (4/11) & \num{-933} (-0.3\%) \\
                      & 162  & 6  & \num{120992}  & \num{244} (0.2\%)    & 14 m 25 s & 5 (2/2)   & \emph{infeasible} \\
                      & 1354 & 5  & \num{1481637} & \num{1605} (0.1\%)   & 12 m 7 s  & 26 (2/23) & \num{1605} (0.1\%) \\
\midrule
\multirow{3}{*}{None} & 30   & 1  & \num{18044}   & \num{0} (0.0\%)      & 1.5 s     & 1 (0/0)   & \num{0} (0.0\%) \\
                      & 39   & 4  & \num{249672}  & \num{0} (0.0\%)      & 4.3 s     & 1 (0/0)   & \num{0} (0.0\%) \\
                      & 57   & 0  & \num{49290}   & \num{0} (0.0\%)      & 4.4 s     & 1 (0/0)   & \num{0} (0.0\%) \\
\bottomrule
\end{tabular}

%% file: Results/case_featured/meta.tex
\def\CaseCzero{242296}
\def\CaseCH{197613}
\def\CaseXmin{1}
\def\CaseXmax{49}
\def\CaseXmaxPre{11}

%% file: Results/cut_comparison_table.tex
\begin{tabular}{l rr rr rr}
\toprule
 & \multicolumn{2}{c}{Our method} & \multicolumn{2}{c}{Benders} & \multicolumn{2}{c}{Combined} \\
\cmidrule(lr){2-3}\cmidrule(lr){4-5}\cmidrule(lr){6-7}
System & Savings (\$, \%) & Time & Savings (\$, \%) & Time & Savings (\$, \%) & Time \\
\midrule
5    & \num{2651} (3.5\%)   & 2.5 s    & \num{2651} (3.5\%)   & 2.7 s     & \num{2651} (3.5\%)   & 3.2 s \\
24   & \num{6107} (5.3\%)   & 38.4 s   & \num{10347} (8.9\%)  & 3 m 46 s  & \num{12248} (10.6\%) & 1 m 18 s \\
73   & \num{3043} (0.9\%)   & 48 m 8 s & \num{0} (0.0\%)      & 12 m 45 s & \num{2496} (0.7\%)   & 36 m 38 s \\
118  & \num{44683} (18.4\%) & 60 m 23 s& \num{17209} (7.1\%)  & 20 m 54 s & \num{17209} (7.1\%)  & 25 m 16 s \\
162  & \num{244} (0.2\%)    & 14 m 25 s& \num{244} (0.2\%)    & 18 m 57 s & \num{244} (0.2\%)    & 25 m 52 s \\
793  & \num{5962} (10.2\%)  & 30 m 21 s& \num{6859} (11.7\%)  & 28 m 31 s & \num{2538} (4.3\%)   & 17 m 12 s \\
1354 & \num{1605} (0.1\%)   & 12 m 7 s & \num{2173} (0.1\%)   & 16 m 0 s  & \num{2173} (0.1\%)   & 14 m 34 s \\
\bottomrule
\end{tabular}

%% file: Sections/Discussion.tex
\section{Discussion}
\label{sec:discussion}

\emph{Planning relaxation and cut choice.} The framework decouples proposal from certification: the planner is a fast candidate generator and the per-step AC check is the sole arbiter of feasibility, so the planner's fidelity affects only how many candidates are screened, not the feasibility of the executed trajectory. A DC planner suffices on the congested conditions studied here and could be replaced by any tighter relaxation, such as a second-order-cone planner~\cite{Heidarifar_2021}, or even a topology sampler, provided it enforces the accumulated cuts. The same decoupling underlies the cut comparison of Section~\ref{subsec:cut_comparison}, where the no-good cuts, each excluding exactly one certified-infeasible topology, prove more reliable than the locally valid Benders duals.

\emph{Computational cost.} Solve times range from seconds to about an hour on the most congested large systems, all on laptop-class hardware and computed offline, so they are compatible with operational planning horizons. The congestion-driven prescreen keeps the switching search tractable, and warm-starting the planner and reusing cuts across operating points are promising ways to reduce solve time.

\emph{Design choices.} The single-action restriction lengthens the transition in exchange for step-by-step interpretability, and it relaxes directly to a bounded number of simultaneous actions within the same MPC structure. The planner minimizes the terminal cost rather than the path-integrated cost, which is justified because a transition executes within minutes while the reconfigured point is held far longer, making the running cost second-order; that running-cost term of Model~\ref{model:gtp} can be reinstated for slower transitions.

\emph{Scope and limitations.} The feasibility certificate is steady-state: intermediate operating points are AC-feasible, but switching transients are not modeled and $N\!-\!1$ security is not enforced along the transition. A target topology can be incorporated through the optional destination term in the objective~\eqref{eq:obj}; because none is used here, savings are measured against the do-nothing ACOPF cost rather than a global NTR optimum, which is intractable to compute at these scales. The candidate substations are fixed upstream by a congestion-based prescreen, which may overlook beneficial splits at buses not adjacent to a congested line, and each system is evaluated at a single operating point, so robustness of the reported savings across load conditions is not established. Finally, the value of a transition depends on congestion: on lightly loaded systems the planner correctly returns near-trivial trajectories (Section~\ref{sec:results}).

%% file: Sections/Conclusion.tex
\section{Conclusion}
\label{sec:conclusion}

We introduced the Optimal Transition Planning (OTP) problem: co-planning the coordinated sequence of substation switching and generation redispatch that carries a grid from its current operating point to a reconfigured one, subject to generator ramp limits, per-step switching budgets, and AC feasibility of every intermediate state. We solved it with a receding-horizon model predictive control (MPC) framework in which a DC planner proposes candidate transitions, a full ACOPF certifies each committed topology, and infeasible topologies are excluded through combinatorial no-good cuts. This keeps the combinatorial search tractable while ensuring that the executed trajectory respects the AC power flow and thermal limits at every step.

Across congested PGLib-OPF systems, the method produces AC-feasible transitions and recovers substantial savings where reconfiguration is valuable, for instance, an $18.4\%$ cost reduction on the 118-bus system through a 49-step transition. The AC filter is decisive: on the most valuable systems an unfiltered DC plan is not AC-realizable at all, and among cut strategies the combinatorial filter proved the most robust. By turning an optimal topology from a static target into an executable, auditable plan on commodity hardware, this work offers a concrete pathway toward network topology reconfiguration (NTR) in operation.

Natural extensions include co-optimizing the destination topology, enforcing $N\!-\!1$ security along the trajectory, adapting the framework for post-contingency recovery so that a certified transition becomes a precomputed response plan, and modeling reactive power, voltage, and switching transients.

%% file: Sections/Appendix.tex
\section{Appendix}

\allowdisplaybreaks

\subsection{Power flow models}
\label{sec:pf}

The network has generators $\mathcal{G}$, buses $\mathcal{N}$, loads $\mathcal{D}$, and branches $\mathcal{L}$; $\mathcal{G}_i$ and $\mathcal{D}_i$ collect the generators and loads at bus $i$, and branch $l$ runs from bus $f_l$ to bus $t_l$. Generator $g$ has active and reactive output $P_g,Q_g$ and cost coefficient $C_g$, bus $i$ has voltage magnitude $V_i$ and angle $\delta_i$, and $\overline{P}_l,\overline{S}_l$ are the active and apparent power ratings of branch $l$.

The DC power flow model is
\begin{subequations} \label{eq:DCPF}
\begin{align}
& P_{g}^{\min} \leq P_{g} \leq P_{g}^{\max}, \quad \forall g \in \mathcal{G} \label{eq:dcpf_Pgen}\\
    & \sum_{g \in \mathcal{G}_{i}} P_{g} - \sum_{d \in \mathcal{D}_{i}} P_d \notag\\
    & \qquad = \sum_{l \in \mathcal{L}|f_{l}=i} P_{l} - \sum_{l \in \mathcal{L}|t_{l}=i} P_{l}, \quad \forall i \in \mathcal{N} \label{eq:dcpf_nodal}\\
    & P_{l} = B_{l}\theta_l, \quad \forall l \in \mathcal{L} \label{eq:dcpf_pf}\\
    & -\overline{P}_{l} \leq P_{l} \leq \overline{P}_{l}, \quad \forall l \in \mathcal{L} \label{eq:dcpf_thermal}
\end{align}
\end{subequations}
and the AC power flow model imposes the generation, voltage, and angle limits ($P_g^{\min}\!\le\! P_g\!\le\! P_g^{\max}$, and likewise $Q_g$, $V_i$, $\delta_i$, with $\delta_{\mathrm{ref}} = 0$) together with
\begin{subequations} \label{eq:ACPF}
\begin{align}
& \sum_{g \in \mathcal{G}_{i}} P_{g} - \sum_{d \in \mathcal{D}_{i}} P_d \notag\\
    & \qquad = \sum_{l \in \mathcal{L}|f_{l}=i} P^{f}_{l} + \sum_{l \in \mathcal{L}|t_{l}=i} P^{t}_{l}, \quad \forall i \in \mathcal{N} \label{eq:acopf_pbal}\\
    & \sum_{g \in \mathcal{G}_{i}} Q_{g} - \sum_{d \in \mathcal{D}_{i}} Q_d \notag\\
    & \qquad = \sum_{l \in \mathcal{L}|f_{l}=i} Q^{f}_{l} + \sum_{l \in \mathcal{L}|t_{l}=i} Q^{t}_{l}, \quad \forall i \in \mathcal{N} \label{eq:acopf_qbal}\\
    & P^{f}_{l} = V_{f_l}^{2}G_{l} \notag\\
    & \quad - V_{f_l}V_{t_l}\big(G_{l}\cos\theta_{l} + B_{l}\sin\theta_{l}\big), \quad \forall l \in \mathcal{L} \label{eq:acopf_Pft}\\
    & Q^{f}_{l} = -V_{f_l}^{2}\!\left(B_{l}+\tfrac{b^{sh}_{l}}{2}\right) \notag\\
    & \quad - V_{f_l}V_{t_l}\big(G_{l}\sin\theta_{l} - B_{l}\cos\theta_{l}\big), \quad \forall l \in \mathcal{L} \label{eq:acopf_Qft}\\
    & (P^{s}_{l})^{2} + (Q^{s}_{l})^{2} \leq \overline{S}_{l}^{2}, \quad \forall l \in \mathcal{L},\ s \in \{f,t\} \label{eq:acopf_thermal}
\end{align}
\end{subequations}
where $\theta_l = \delta_{f_l}-\delta_{t_l}$, and the to-side flows $P^{t}_{l}$ and $Q^{t}_{l}$ are given by the same expressions with the endpoints $f_l$ and $t_l$ exchanged (equivalently $\theta_l \to -\theta_l$). Here $G_l + jB_l = 1/(r_l + jx_l)$ is the series admittance and $b^{sh}_l$ the line charging susceptance.

\subsection{Augmented Network Representation}
\label{sec:switching}

\begin{figure}[tbp]
    \centering
    \includegraphics[scale = 0.44]{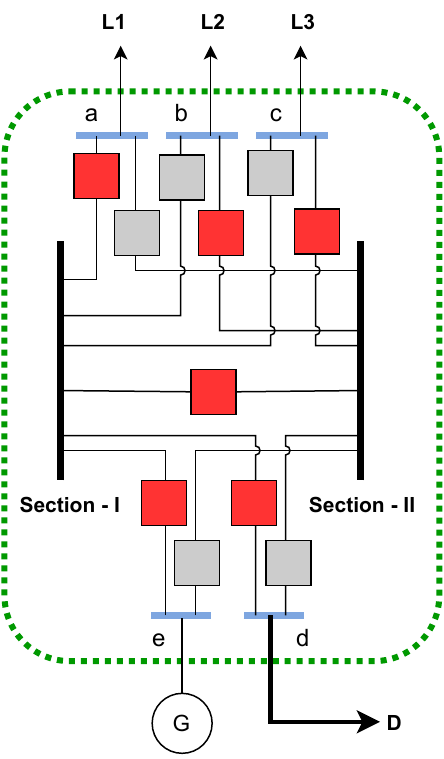}
    \caption{Example of augmented network representation of a substation (enclosed by a dashed rectangle). \protect \includegraphics[scale=0.4]{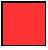} switched on auxiliary line \protect \includegraphics[scale=0.4]{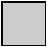} switched off auxiliary line \protect \includegraphics[scale=0.4]{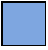} auxiliary bus. L represents a transmission line, D for demand, and G for generator.}
    \label{fig:ANR}
\end{figure}

We model substation switching with the augmented network representation (ANR)~\cite{morsy_security_2022}, illustrated in Figure~\ref{fig:ANR}: each grid element gets a dedicated auxiliary bus, the two bus-bar sections are linked by a switchable coupler, and each auxiliary bus connects to both sections through two mutually exclusive switches. Let $\mathcal{N}^{aux}$ denote the auxiliary buses, $\mathcal{S}$ the splittable substations, and $\mathcal{E}^{r},\mathcal{E}^{c}$ the reconfiguration and coupler switches; $\mathcal{E}^{r}_{i} \subseteq \mathcal{E}^{r}$ are the two reconfiguration switches of auxiliary bus $i$. For a substation $s \in \mathcal{S}$, $c_{s}$ is its coupler switch, $b \in \{1,2\}$ indexes its two bus-bar sections, $\mathcal{E}^{r}_{s,b}$ collects the reconfiguration switches joining section $b$, and $(\alpha^{1}_{s}, \alpha^{2}_{s})$ are the two legs of a designated anchor element, with $\alpha^{1}_{s}$ to section~1. Applying a switch pattern $\tau$ yields the \emph{reduced network}, obtained by opening the de-energized auxiliary branches and merging any sections joined by a closed coupler.

The resulting switching constraints are
\begin{subequations} \label{eq:sss}
\begin{align}
 & z^{r}_{l} \in \{0,1\}\ \ \forall l \in \mathcal{E}^{r}, \quad z^{c}_{l} \in \{0,1\}\ \ \forall l \in \mathcal{E}^{c} \label{eq:switch_bin}\\
    & \sum_{l \in \mathcal{E}^{r}_{i}} z^{r}_{l} = 1, \quad \forall i \in \mathcal{N}^{aux} \label{eq:mutual_exclusivity}\\
    & \lvert \delta_{f_{l}} - \delta_{t_{l}} \rvert \leq (1-z^{r}_{l})M_{l}^{\delta}, \quad \forall l \in \mathcal{E}^{r} \label{eq:reconf_angle_equality}\\
    & \lvert P_{l} \rvert \leq z^{r}_{l}M^{\mathcal{E}}_l, \quad \forall l \in \mathcal{E}^{r} \label{eq:reconf_line_capacity}\\
    & \lvert \delta_{f_{l}} - \delta_{t_{l}} \rvert \leq (1-z^{c}_{l})M_{l}^{\delta}, \quad \forall l \in \mathcal{E}^{c} \label{eq:split_angle_equality}\\
    & \lvert P_{l} \rvert \leq z^{c}_{l}M^{\mathcal{E}}_l, \quad \forall l \in \mathcal{E}^{c} \label{eq:splitting_line_capacity}\\
    & z^{r}_{\alpha^{1}_{s}} \geq z^{r}_{\alpha^{2}_{s}}, \quad \forall s \in \mathcal{S} \label{eq:lex}\\
    & z^{r}_{l} \geq z^{c}_{c_{s}}, \quad \forall l \in \mathcal{E}^{r}_{s,1},\ \forall s \in \mathcal{S} \label{eq:coupler_pin}\\
    & \sum_{l \in \mathcal{E}^{r}_{s,b}} z^{r}_{l} \geq 2\,(1 - z^{c}_{c_{s}}), \quad \forall b \in \{1,2\},\ \forall s \in \mathcal{S} \label{eq:min_load}
\end{align}
\end{subequations}
in which the reconfiguration switches $z^{r}$ and coupler switches $z^{c}$ are big-M gated, the routing constraint~\eqref{eq:mutual_exclusivity} assigns each auxiliary bus to exactly one bus-bar section, and~\eqref{eq:lex}--\eqref{eq:min_load} are valid inequalities that break switching symmetries. The constants $M_{l}^{\delta}$ and $M^{\mathcal{E}}_{l}$ are big-M values bounding the angle difference and flow of a de-energized branch; we use $M_{l}^{\delta} = 1.2$~rad and $M^{\mathcal{E}}_{l} = 50\,S_{\mathrm{base}}$. The flow gates~\eqref{eq:reconf_line_capacity} and~\eqref{eq:splitting_line_capacity} are written for the DC line flow $P_{l}$; under AC physics they apply to both branch-end flows, active and reactive, $\lvert P^{f}_{l}\rvert, \lvert P^{t}_{l}\rvert, \lvert Q^{f}_{l}\rvert, \lvert Q^{t}_{l}\rvert \leq z^{r}_{l} M^{\mathcal{E}}_{l}$ (and analogously with $z^{c}_{l}$).

\subsection{Generalized-Benders baseline}
\label{sec:benders}

As a literature-derived comparator we implement the AC-Benders strategy of~\cite{Nasrolahpour_2012}, originally a single-snapshot scheme, generalized here to the multi-period transition planner. The MPC-DC planner (Section~\ref{subsec:mpcx}) serves as the master and proposes an entire $H$-step trajectory, while the multi-period ACOPF (Model~\ref{model:mpacopf}), solved with the switches fixed to that proposal, serves as the subproblem; its duals across all steps form generalized-Benders cuts on an auxiliary recourse term (weight $\rho$) added to the master, namely a feasibility cut when the trajectory is AC-infeasible and an optimality cut when it is feasible but costlier than the incumbent, as constructed in~\cite{Nasrolahpour_2012}. Because the AC subproblem is non-convex these duals are only local, so the cuts are heuristic and are discarded across receding steps, whereas the no-good cuts of Section~\ref{subsec:filter} persist; Section~\ref{subsec:cut_comparison} reports the comparison.

\subsection{Full Algorithm}
\label{sec:algorithm}

Algorithm~\ref{alg:mpc} states the complete receding-horizon procedure as implemented; Figure~\ref{fig:method_overview} gives its high-level structure.

\begin{algorithm}[t]
\caption{Receding-horizon MPC-DC with AC filtering}
\label{alg:mpc}
\begin{algorithmic}[1]
\REQUIRE grid, optional target topology, horizon $H$, switching budgets $\lambda_r,\lambda_c$, tolerances $\epsilon$ and drift
\STATE Solve DC and AC OPF at the initial topology $\tau_0$; set the starting operating point and the cost bounds
\STATE Initialize an empty topology-cut store
\WHILE{not terminated}
    \STATE Solve MPC-DC (Model~\ref{model:mpcdc}) with all stored cuts to obtain a planned $H$-step trajectory
    \STATE Screen each unique planned topology with the single-period AC proxy
    \IF{some topology fails the proxy}
        \STATE add AC-infeasibility cut(s) and re-plan
    \ELSE
        \STATE Solve the multi-period ACOPF (Model~\ref{model:mpacopf}) over the planned sequence
        \IF{infeasible}
            \STATE add partial-trajectory cut(s) and re-plan
        \ELSIF{terminal cost not improved}
            \STATE add a full-trajectory cut and re-plan
        \ELSE
            \STATE accept the plan; commit the first action; record AC-feasible topologies
            \IF{committed topology unchanged and improvement $<\epsilon$}
                \STATE terminate
            \ENDIF
        \ENDIF
    \ENDIF
\ENDWHILE
\STATE Expand the committed trajectory to restore ramp feasibility
\RETURN AC-feasible, ramp-feasible transition plan
\end{algorithmic}
\end{algorithm}